%% file: CameraReady2027.tex
\documentclass[letterpaper]{article} 
\usepackage{aaai2027}  
\usepackage[hyphens]{url}  
\usepackage{graphicx} 
\usepackage{natbib}  
\usepackage{caption} 
\usepackage{algorithm}
\usepackage{algorithmic}
\usepackage{subcaption} 
\usepackage{newfloat}
\usepackage{xspace}
\usepackage{listings}
\DeclareCaptionStyle{ruled}{labelfont=normalfont,labelsep=colon,strut=off} 
\floatstyle{ruled}
\newfloat{listing}{tb}{lst}{}
\floatname{listing}{Listing}

\usepackage{booktabs}
\newcommand{\name}{\textit{HypoForge}\xspace}

\nocopyright

\title{\name: A Self-Improving Multi-Agent Framework for Automated Hypothesis Generation and Testing via Scientific Skill Learning}

\author{
    Ziqing Qian\textsuperscript{\rm 1,\rm 2},
    Jiaying Lei\textsuperscript{\rm 1,\rm 2,\rm 3}, 
    Yifang Wang\textsuperscript{\rm 4},
    Nan Cao\textsuperscript{\rm 1,\rm 2,\rm 3}\thanks{Corresponding author.}
}
\affiliations{
    \textsuperscript{\rm 1}Intelligent Big Data Visualization Lab, Tongji University, Shanghai, China\\
    \textsuperscript{\rm 2}Shanghai Research Institute for Intelligent Autonomous System, Tongji University, Shanghai, China\\
    \textsuperscript{\rm 3}Shanghai Innovation Institute, Shanghai, China\\
    \textsuperscript{\rm 4}Florida State University Tallahassee, Florida, USA\\
    
    2411920@tongji.edu.cn,
    jiaying.lei@outlook.com,
    yifang.wang@fsu.edu,
    nan.cao@gmail.com
}

\begin{document}

\maketitle

\input{Sections/00abstract}


\input{Sections/01introduction}

\input{Sections/02related_work}

\input{Sections/03method}
\input{Sections/04experiment}

\input{Sections/05discussion}

\input{Sections/06conclusion}

\newpage
\bibliography{aaai2027}


\end{document}

%% file: Sections/00abstract.tex
\begin{abstract}
Large language models (LLMs) have enabled AI scientist systems to automate scientific discovery, yet existing approaches most rely on static prompting or fixed workflows and fail to accumulate experience for continual improvement. We propose \textbf{\name}, an experience-guided multi-agent framework that learns reusable scientific skills for automated hypothesis generation and hypothesis testing.
\name is built on the observation that these two stages involve different supervision signals. For hypothesis generation, where explicit feedback is unavailable, \name adopts an adversarial generator--discriminator mechanism to improve reasoning through comparative critique. For hypothesis testing, where empirical feedback is available, \name learns testing skills from execution outcomes and ground-truth results. By matching skill learning strategies with stage-specific supervision, \name enables continual improvement without fine-tuning foundation models.
Experiments on hypothesis generation and testing benchmarks show that \name consistently outperforms existing AI scientist frameworks and skill-level variants. Further analysis demonstrates the effectiveness of the proposed stage-specific skill learning
paradigms. 
\end{abstract}

%% file: Sections/01introduction.tex
\section{Introduction}
Hypothesis generation and hypothesis testing constitute the core of hypothesis-driven scientific research. Given a research question and empirical observations, scientists formulate candidate hypotheses to explain underlying phenomena and subsequently validate them through statistical analyses and experimental evidence. More importantly, scientific research is inherently cumulative: researchers continuously summarize reusable scientific skills from previous successes and failures, enabling them to generate better hypotheses, design more reliable experiments, and solve increasingly complex scientific problems.

Recent advances in large language models (LLMs) have led to the emergence of AI scientist systems capable of automating scientific reasoning~\cite{wang2023scientific}, hypothesis generation~\cite{zhou2024hypothesis}, experiment design~\cite{boiko2023autonomous}, code generation~\cite{yang2024swe}, and data analysis~\cite{gu2024blade}. Although these systems have demonstrated impressive reasoning capabilities, most either focus on a single stage of scientific discovery, producing hypotheses that are never tested against data or testing hypotheses that must be supplied externally, or operate through static prompting or predefined workflows, limiting their ability to distill, accumulate, and reuse scientific experience across tasks. As a result, they repeatedly pay similar reasoning efforts from scratch across tasks, leading to reduced discovery efficiency~\cite{wang2023voyager}.

We therefore ask how the discovery experience can be distilled into reusable scientific skills across the two stages that constitute the discovery loop: hypothesis generation and hypothesis testing.
The central challenge is that the two stages differ fundamentally in the supervision available for skill learning.
Hypothesis generation is an open-ended scientific reasoning task that lacks explicit supervision, making it difficult to directly assess hypothesis quality. Consequently, improvement relies on comparing and critiquing alternative hypotheses to gradually distill effective reasoning strategies. In contrast, hypothesis testing naturally provides empirical supervision through executable experiments and ground-truth outcomes, enabling direct evaluation of experimental design and implementation. 
These fundamentally different supervision signals imply that hypothesis generation and hypothesis testing require stage-specific learning paradigms rather than a unified refinement strategy.

To address this, we propose \textbf{\name}, a self-improving multi-agent framework for automated hypothesis generation and hypothesis testing. Rather than simply storing previous reasoning trajectories, \name continuously distills accumulated experience into reusable scientific skills that improve future reasoning. Specifically, for hypothesis generation, \name introduces an adversarial generator--discriminator framework that iteratively learns hypothesis-generation skills through comparative critique without explicit supervision. For hypothesis testing, HypoForge exploits execution outcomes and ground-truth feedback to iteratively learn testing skills under empirical supervision. By aligning skill learning with the supervision characteristics of each stage, \name enables continual capability improvement without requiring fine-tuning of the underlying foundation models.
We evaluate \name on benchmark datasets for automated hypothesis generation and hypothesis testing. Experimental results demonstrate that \name consistently outperforms existing AI scientist frameworks and skill-level variants in both hypothesis quality and testing performance. Extensive ablation studies further verify the effectiveness of the proposed stage-specific skill learning paradigms. 
Our contributions are summarized as follows:

\begin{itemize}

\item We propose \name, a self-improving multi-agent framework that learns reusable scientific skills for automated hypothesis generation and hypothesis testing.

\item We identify the distinct supervision characteristics of hypothesis generation and hypothesis testing, and develop two stage-specific skill learning paradigms: an adversarial self-improvement mechanism for hypothesis generation without explicit supervision, and a ground-truth feedback learning mechanism for hypothesis testing with empirical supervision.

\item Extensive experiments demonstrate that \name consistently improves both hypothesis generation and hypothesis testing over existing AI scientist frameworks, while the learned skills exhibit strong transferability across diverse scientific research tasks.

\end{itemize}

%% file: Sections/02related_work.tex
\section{Related Work}
In this section, we review two research directions most related to our work: autonomous scientific discovery and skill learning for LLM agents. 

\subsection{Autonomous Scientific Discovery}

Autonomous scientific discovery aims to enable machines to generate and validate novel scientific hypotheses from empirical evidence~\cite{wang2023scientific}. 
Central to autonomous scientific discovery are hypothesis generation and hypothesis testing, which form an iterative reasoning loop that transforms empirical observations into testable explanations and evaluates whether those explanations are supported by evidence.
With recent advances in artificial intelligence (AI) and large-scale data-driven approaches, automated hypothesis generation and testing have attracted increasing attention. Takagi et al. demonstrated the potential of LLMs for scientific ideation through knowledge synthesis \cite{takagi2023towards}, while Zhou et al. showed that prompting strategies and contextual information significantly affect generated hypotheses \cite{zhou2024hypothesis}. Recent benchmarks, including HypoBench and DiscoveryBench, have established standardized evaluations for LLM-based hypothesis generation \cite{liu2025hypobench,majumder2025discoverybench}. Beyond generation tasks, a few studies have also explored automated hypothesis testing by integrating LLM reasoning with computational experiments and iterative testing, including self-improving scientific agents \cite{brunnsaaker2025self}, sequential falsification frameworks \cite{huang2025automated}, and multi-agent discovery workflows \cite{gupta2026accelerating}.

Despite these advances, existing systems typically focus on a single stage of scientific discovery, either hypothesis generation or hypothesis testing, and rely on the pretrained model's fixed reasoning ability to solve each new task, thus limiting their ability to accumulate experience and progressively improve their scientific reasoning. 
Our work bridges this gap through agent skill learning, integrating hypothesis generation and testing into a unified discovery process while distilling accumulated discovery experiences into reusable reasoning capabilities.

\subsection{Agent Skill Learning}
Recent LLM agents have evolved from stateless reasoning toward experience-driven capability improvement. Memory-augmented agents, such as Generative Agents and MemGPT, introduced mechanisms to store and retrieve long-term information to support future interactions \cite{park2023generative,packer2023memgpt}. 
Reflexion further enable agents to learn from past failures \cite{shinn2023reflexion}, while Self-Refine demonstrated iterative refinement through self-generated feedback without updating model parameters \cite{madaan2023self}.
Building upon these ideas, a few studies have investigated agent skill learning as a way to distill reusable capabilities from accumulated past experiences. Voyager pioneered automatic skill library construction, allowing agents to discover, store, and reuse executable skills during open-ended interaction \cite{wang2023voyager}. Subsequent research has extended this paradigm from skill storage toward skill acquisition, optimization, and refinement. For example, reinforcement learning-based approaches explore how agents can improve skill libraries through interaction experiences \cite{wang2026reinforcement}, while another line of research investigates automatic skill optimization and revision through trajectory feedback and testing approaches \cite{yang2026skillopt,liu2026skillrevise}. More recent studies further explore self-evolving skill frameworks that improve agent capabilities through iterative skill generation, evaluation, and refinement\cite{zhang2026coevoskills,vishe2026skill}.

However, existing skill-learning approaches typically optimize agent skills in a unified manner, aggregating experiences and feedback from different stages without explicitly modeling their distinct capabilities. 
This limits stage-specific skill refinement and feedback attribution, making it difficult to determine how accumulated experiences should be transformed into targeted procedural skills for complex multi-step workflows.
In this work, we address this problem by designing a stage-specific skill learning framework that leverages different feedback to distill transferable strategies for hypothesis generation and testing.

%% file: Sections/03method.tex
\begin{figure*}[t]
    \centering
    \includegraphics[width=\textwidth]{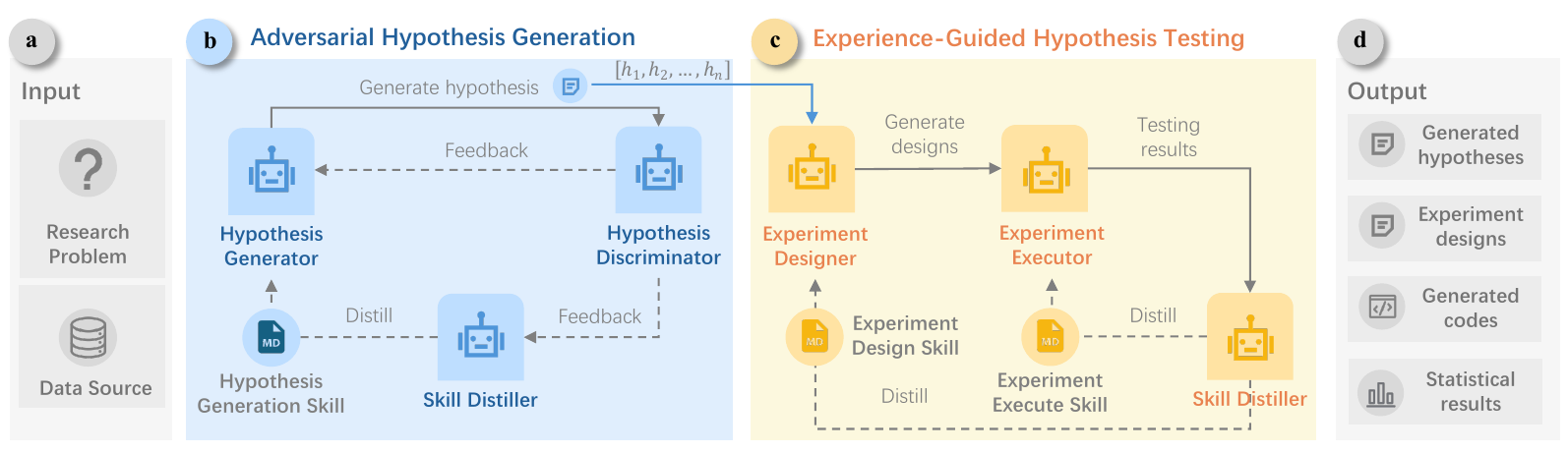}
    \caption{Multi-agent framework that supports hypothesis generation and testing with skill learning. }
    \label{fig:framework}
\end{figure*}

\section{Problem Formulation}

Given a user research problem $P$ and an associated data source $D$, the multi-agent system aims to generate a set of hypotheses $\mathcal{H}=\{h_i\}_{i=1}^{M}$ and corresponding testing trajectories $\mathcal{V}=\{\tau_i\}_{i=1}^{M}$, where $\tau_i=(h_i,e_i,c_i,r_i)$ denotes the testing trajectory of each hypothesis, including the hypothesis ($h_i$), experimental design ($e_i$), executable code ($c_i$), and execution outcome ($r_i$).
For each iteration $t$, the system generates hypotheses $\mathcal{H}^{t}$ and testing trajectories $\mathcal{V}^{t}$ based on the current skills and accumulated discovery experiences. 
The objective is to progressively learn three reusable procedural agent skills $\mathcal{S}^{t}=\{\mathcal{S}_{h}^{t},\mathcal{S}_{e}^{t},\mathcal{S}_{x}^{t}\}$ for hypothesis generation, experimental design, and experiment execution, respectively. 
The skills are iteratively updated by the multi-agent system based on newly acquired experiences and feedback from training records:
\begin{equation}
\mathcal{S}_{h}^{t+1}
=\Phi_{h}(\mathcal{S}_{h}^{t},\mathcal{H}^{t},\mathcal{F}_{h}^{t}),
\label{eq:skill_h}
\end{equation}
\begin{equation}
\mathcal{S}_{e}^{t+1},\mathcal{S}_{x}^{t+1}
=\Phi_{t}(\mathcal{S}_{e}^{t},\mathcal{S}_{x}^{t},\mathcal{V}^{t}),
\label{eq:skill_v}
\end{equation}
where $\Phi_{h}$ denote the skill learning processes for hypothesis generation, $\mathcal{F}_{h}^{t}$ denotes the feedback obtained from hypothesis evaluation, and $\Phi_{t}$ for hypothesis testing.

\section{Method}
In this section, we present {\name}, a stage-wise skill learning framework for scientific hypothesis generation and testing. As illustrated in Fig.~\ref{fig:framework}, the framework decomposes hypothesis-driven scientific research process into two stages and learns dedicated skills for each stage according to its distinct reasoning process and supervision signals. The adversarial hypothesis generation stage (Fig.~\ref{fig:framework}(b)) iteratively refines hypothesis construction through interactions between a hypothesis generator and a discriminator, distilling multi-dimensional evaluation feedback into reusable hypothesis-generation skills. The experience-guided hypothesis testing stage (Fig.~\ref{fig:framework}(c)) learns reusable testing skills from prior testing trajectories by exploiting ground-truth outcomes and feedback from experiment design, execution, and validation.

\subsection{Adversarial Hypothesis Generation}
Inspired by adversarial learning in Generative Adversarial Networks (GANs)~\cite{goodfellow2014generative}, we formulate hypothesis generation as an adversarial skill learning problem (Fig.~\ref{fig:framework}(b)). Specifically, a hypothesis generator produces a batch of candidate hypotheses that approximates the distribution of plausible scientific discoveries, while a multi-dimensional discriminator evaluates the generated hypothesis set as a whole. Unlike the binary discriminator in conventional GANs, our discriminator is formulated as a scoring function that measures how closely the generated hypothesis distribution matches the characteristics of high-quality human-written scientific hypotheses. The resulting distribution-level feedback is distilled into reusable hypothesis-generation skills, enabling the generator to progressively improve its generation policy across iterative interactions. Compared with Reviewer/Critic-based multi-agent frameworks, which iteratively revise individual generation items using instance-level feedback, our framework performs distribution-level optimization over the entire hypothesis set. Consequently, the learned skills capture transferable generation strategies, including variable selection, causal reasoning, hypothesis construction, and avoidance of common hypothesis generation failures, rather than hypothesis-specific corrections, leading to better generalization across different hypothesis-driven scientific research tasks.

\subsubsection{Hypothesis Generator}

Given a research problem $P$ and data source $D$, the hypothesis generator $G_h$ produces a batch of candidate hypotheses,

\begin{equation}
\mathcal{H}^{t}
=
\{h_1^{t},h_2^{t},\ldots,h_N^{t}\}
=
G_h(P,D,\mathcal{S}_{h}^{t}),
\end{equation}
where $\mathcal{S}_{h}^{t}$ denotes the hypothesis generation skill accumulated from previous adversarial interactions.

The generated hypothesis set provides a distributional approximation of plausible scientific discoveries, enabling the discriminator to jointly evaluate the scientific quality, diversity, and discovery coverage of the generated hypotheses.
Each hypothesis is represented as a structured scientific claim describing a testable relationship between variables together with its expected direction. The generator prompt specifies only the agent role, research objective, output format, and general constraints, while the learned skill $\mathcal{S}_{h}^{t}$ encapsulates reusable procedural knowledge distilled from previous adversarial interactions, including informative variable selection, causal reasoning patterns, hypothesis construction strategies, and common failure avoidance.

\subsubsection{Multi-Dimensional Hypothesis Discriminator}

The proposed discriminator $D_h$ is formulated as a multi-dimensional scoring function rather than a binary classifier. Instead of determining whether a generated hypothesis is ``real'' or ``fake'', $D_h$ estimates how closely the generated hypothesis set matches the latent distribution of high-quality human-written scientific hypotheses.

High-quality scientific hypotheses generally satisfy several fundamental principles established in the scientific discovery literature, including empirical grounding~\cite{king2009automation,langley1987scientific}, causal validity~\cite{neuberg2003causality}, falsifiability~\cite{popper2005logic}, theoretical consistency~\cite{lakatos1978methodology}, clarity and operationalizability~\cite{shapere1964structure}, and scientific novelty~\cite{fortunato2018science}. These principles are summarized into seven evaluation dimensions:
\begin{equation}
\mathbf d_i
=
[D_1^i,D_2^i,\ldots,D_7^i],
\end{equation}
where
$D_1,D_2,D_5,D_6\in[0,1]$
measure empirical grounding, causal plausibility, structural clarity, and scientific novelty, respectively, while
$D_3,D_4,D_7\in\{0,1\}$
serve as hard constraints enforcing falsifiability, theoretical consistency, and empirical verifiability.  Collectively, these measurements (implemented based LLM detailed in the supplement materials) characterize the essential properties shared by high-quality human-written scientific hypotheses as follows:
\begin{equation}
s_i
=
D_3^iD_4^iD_7^i
\left(
D_1^iD_2^iD_5^iD_6^i
\right)^{\frac14},
\end{equation}
where larger values indicate greater consistency with the characteristics of high-quality human-written hypotheses. The overall scientific quality of the generated hypothesis set is computed as:
\begin{equation}
Q(\mathcal H^t)
=
\frac1N
\sum_{i=1}^{N}
s_i.
\end{equation}

Besides evaluating hypothesis quality, the discriminator further compares the generated hypothesis set with reference hypotheses to provide reflective feedback:
\begin{equation}
C(\mathcal H^t)
=
Match(\mathcal H^t,\mathcal H_{gt}),
\end{equation}
where $\mathcal H_{gt}$ denotes the reference hypothesis set. 
Unlike hypothesis quality, this comparison provides additional insights into the discovery patterns covered by the generated hypotheses. 
Specifically, hypotheses with high overlap with reference hypotheses reveal effective scientific patterns, while hypotheses with limited overlap indicate potential missing factors or unexplored directions for future exploration.




\subsubsection{Adversarial Skill Learning}

Inspired by adversarial learning, the generator progressively improves its hypothesis generation policy by maximizing the scientific quality score,

\begin{equation}
\max_{G_h}
\mathcal L_G
=
Q(\mathcal H^t).
\end{equation}

Unlike conventional GANs, the discriminator is not optimized through binary classification, nor is the generator updated via gradient back-propagation. Instead, the discriminator summarizes the strengths and weaknesses of the generated hypothesis set and distills them into reusable procedural knowledge:
\begin{equation}
\mathcal S_h^{t+1}
=
Distill
\left(
\mathcal S_h^{t},
\mathcal H^{t},
Q(\mathcal H^{t}),
C(\mathcal H^{t})
\right),
\end{equation}
where $Distill(\cdot)$ is a LLM-based agent that extracts transferable hypothesis-generation strategies from discriminator feedback and accumulated hypothesis experiences.

The updated skill $\mathcal S_h^{t+1}$ is incorporated into the generator in the next iteration, enabling continuous improvement of the generation policy. By distilling distribution-level feedback into reusable procedural knowledge, the generator progressively aligns with the distribution of high-quality human-written hypotheses, achieving stronger generalization across hypothesis-driven scientific research tasks.


\subsection{Experience-Guided Hypothesis Testing}

Hypothesis testing systematically evaluates candidate scientific hypotheses by transforming them into executable empirical studies and determining whether they are supported by observational evidence. Given a hypothesis, the testing process consists of designing an experiment protocol, implementing the corresponding testing procedure, executing the analysis on the target dataset, and interpreting the resulting evidence. Unlike hypothesis generation, hypothesis testing provides explicit supervision through ground-truth outcomes, allowing the system to evaluate the quality of the entire testing process. Building upon this property, we propose an experience-guided hypothesis testing framework that continuously improves experiment design and execution through iterative skill learning. Specifically, the framework first generates a textual experiment protocol for each hypothesis, then implements and executes the corresponding testing program, and finally compares the obtained testing results with the ground truth to analyze the complete testing trajectory. The resulting recommendations are distilled into reusable experiment design and execution skills, enabling progressively more rigorous and reliable hypothesis testing.
Specifically, the experiment design skill accumulates methodological experience, such as data variable selection, confounder control, and analysis method selection, for constructing reliable experiment plans. 
The execution skill distills engineering experience to realize these experiment plans in code, including code generation, library usage, debugging, and reliable execution of statistical procedures.

\subsubsection{Experiment Design}

Given a hypothesis $h_i$, the experiment designer agent $G_e$ constructs a textual experiment protocol describing how the hypothesis should be empirically tested on the target dataset $D$:
\begin{equation}
e_i=G_e(h_i,D,\mathcal{S}_{e}^{t}),
\end{equation}
where $\mathcal{S}_{e}^{t}$ denotes the learned experiment design skill. The protocol $e_i$ serves as a high-level testing specification that defines all essential components of the experiment, including hypothesis operationalization, variable selection, data preprocessing and filtering, confounder control, statistical analysis methods, significance criteria, evaluation metrics, and the complete experimental workflow. Rather than encoding task-specific instructions, the design skill $\mathcal{S}_{e}^{t}$ provides reusable procedural knowledge distilled from previous testing experiences, enabling the agent to progressively construct more rigorous and scientifically sound experiment protocols.

\subsubsection{Experiment Execution}

Given $e_i$, the execution agent $G_x$ generates an executable testing program $c_i$ as follows:
\begin{equation}
c_i=G_x(e_i,\mathcal{S}_{x}^{t}),
\end{equation}
where $\mathcal{S}_{x}^{t}$ denotes the learned execution skill. $c_i$ is subsequently executed on the target dataset:
\begin{equation}
r_i=Exec(c_i,D),
\end{equation}
where $r_i$ contains the complete testing outcomes, including statistical results, significance tests, estimated effect sizes, and the final testing conclusion.

The execution skill $\mathcal{S}_{x}^{t}$ captures reusable implementation knowledge for translating $e_i$ into reliable programs $c_i$, including statistical implementation, code organization, execution robustness, exception handling, debugging strategies, and standardized result reporting. The complete testing trajectory is then represented as:
\begin{equation}
\tau_i=(h_i,e_i,c_i,r_i),
\end{equation}
which records the hypothesis $h_i$, experiment protocol $e_i$, implementation $c_i$, and testing outcomes $r_i$ for subsequent skill refinement.

\subsubsection{Experience-Based Skill Refinement}
After executing the testing procedure, the framework compares the obtained testing outcome with the corresponding ground-truth result to evaluate the complete testing trajectory. The ground-truth outcome provides explicit supervision for evaluating both the testing conclusion and the validity of the underlying experiment design and implementation.

Specifically, given a testing trajectory $\tau_i=(h_i,e_i,c_i,r_i)$, the framework optimizes the experiment design and execution skills according to the test pass rate:
\begin{equation}
(\mathcal{S}_{e}^{*},\mathcal{S}_{x}^{*})
=
\arg\max_{\mathcal{S}_{e},\mathcal{S}_{x}}
T_h,
\end{equation}
where $T_h$ denotes the proportion of ground-truth hypotheses that are successfully tested by the generated experiments. Through iterative skill refinement, the framework aims to improve the reliability of experiment design and execution for future hypothesis testing.

The accumulated testing trajectories and corresponding ground-truth outcomes are subsequently provided to the skill distiller for experience-based refinement:
\begin{equation}
(\mathcal{S}_{e}^{t+1},\mathcal{S}_{x}^{t+1})
=
Distill
\left(
\mathcal{S}_{e}^{t},
\mathcal{S}_{x}^{t},
\{\tau_i\}_{i=1}^{N}
\right),
\end{equation}
where $Distill(\cdot)$ is an LLM-based agent that analyzes previous testing trajectories and execution outcomes, attributes testing successes and failures to different stages of the experimental process, and extracts transferable experimental practices rather than memorizing individual cases.

The refined skills are incorporated into the experiment designer and execution agents in subsequent iterations, enabling stage-wise improvement of the entire hypothesis testing pipeline. By continuously refining experiment design and execution skills through accumulated testing experiences, the proposed framework acquires transferable testing capabilities that generalize across diverse hypothesis-driven scientific research tasks.

%% file: Sections/04experiment.tex
\section{Experiments}
In this section, we conduct comprehensive experiments to evaluate the effectiveness of our proposed framework. We begin by describing the experimental settings. Then, we compare our framework with baselines on both hypothesis generation and hypothesis testing tasks. Finally, we perform detailed skill learning analysis to investigate skill evolution during refinement, and conduct ablation studies to evaluate the role of different refinement signals.

\subsection{Experimental Setup}
We first introduce the experimental setup, including the benchmark datasets, comparison baselines, evaluation metrics, and implementation details.

\subsubsection{Datasets} 
We evaluate our framework on \textbf{HypoBench}~\cite{liu2025hypobench}, a benchmark for scientific hypothesis generation and discovery. 
The original benchmark contains 13 tasks from diverse domains, and we use all tasks to construct our experimental suite. 
The detailed task adaptation and preprocessing procedures are provided in the supplement materials, and the statistics of all tasks are summarized in Table~\ref{tab:dataset_statistics}.
Each task consists of a research question, an associated dataset, and a set of reference hypotheses. 
For the hypothesis generation stage, we use representative data from the selected 13 tasks. 
For the hypothesis testing stage, we retain only tasks with empirically evaluable ground-truth hypotheses, as trajectory-level supervision from testing outcomes is required to refine experiment design and execution skills.

\begin{table}[htbp]
\centering
\small
\begin{tabular}{lcc}
\toprule
Task & Type & \#Data Size \\
\midrule
Deceptive Reviews Detection & Real & 800 \\
Dreddit Mental Stress Detection & Real & 200 \\
AI-Generated Content Detection & Real & 200 \\
News Headline Engagement & Real & 200 \\
Persuasive Argument Prediction & Real & 200 \\
Retweet Prediction & Real & 200 \\
Paper Citations (Health) & Real & 244 \\
Paper Citations (NeurIPS) & Real & 308 \\
Paper Citations (Radiology) & Real & 392 \\
Shoe Sales & Synthetic & 900 \\
College Admission & Synthetic & 200 \\
Presidential Election & Synthetic & 1,750 \\
Personality Prediction & Synthetic & 1,750 \\
\bottomrule
\end{tabular}
\caption{HypoBench tasks used in our experiments.}
\label{tab:dataset_statistics}
\end{table}

\subsubsection{Baselines}
We compare {\name} against two types of baselines: system-level scientific discovery approaches and skill-level variants.

\begin{itemize}
    \item \textbf{System-level comparison.}
    For \underline{hypothesis generation}, we use HypoGeniC~\cite{zhou2024hypothesis}, which provides an automated pipeline for hypothesis generation, and HypotheSAEs~\cite{movva2025sparse}, which derives hypotheses by interpreting sparse autoencoder features from neural representations.
    For \underline{hypothesis testing}, we include POPPER~\cite{huang2025automated}, a scientific discovery framework that supports automated hypothesis validation, and ReAct~\cite{yao2022react}, which enables LLM agents to iteratively reason, plan, and execute actions for experiment design and validation.
    In addition, we include LLM-ZeroShot and LLM-FewShot~\cite{brown2020language} as general LLM-based baselines for both tasks.
    
    \item \textbf{Skill-level comparison.} 
    We include three skill variants to compare the effectiveness of learned skills on both tasks. 
    (1) No-Skill: removes the learned skill and uses only the original agent framework;
    (2) AI-Generated-Skill: replaces the learned skill with an LLM-generated procedural skill, without refinement through task-specific experiences or evaluation feedback;
    (3) Human-Designed-Skill: replaces the learned skill with a pre-defined manually specified skill without refinement.~\cite{scientific_agent_skills_astropy_2026}
\end{itemize}

\subsubsection{Metrics}
For \underline{hypothesis generation}, we use $Q(\mathcal H^t)$ score to evaluate the quality of generated hypotheses, and $Hit@K$ to measure the coverage of generated hypotheses against reference hypotheses. For \underline{hypothesis testing}, we evaluate the effectiveness of automatically generated experiments using Test Pass Rate $T_h$ and Execution Success Rate $E_h$. 
$T_h$ measures the proportion of ground-truth hypotheses that are successfully validated by the generated experiments, where successful validation requires the experiment to execute successfully and achieve statistical significance ($p<0.05$). 
$E_h$ measures the proportion of experiments that successfully execute and produce valid results.

\subsubsection{Implementation Details}
We divide the selected HypoBench tasks into training and testing sets at a 1:1 ratio, where the training split is used for skill learning and the unseen test split is reserved for evaluation. 
All agents are instantiated with DeepSeek-V4-Flash ~\cite{deepseekai2026deepseekv4} as the backbone model. 
During training, skill updates are retained only when the corresponding learning objective improves; otherwise, the previous skill is preserved for subsequent iterations, and the training process terminates when no improvement is observed for $k$ consecutive attempts ($k=3$).
All reported results are averaged over three independent runs.

\subsection{Hypothesis Generation Evaluation}
We first evaluate the effectiveness of the hypothesis generation stage on test tasks. 
All methods receive the same input (research problem and data source pairs), and their generated hypotheses are evaluated using the same discriminator and reference hypotheses. 
We compare {\name} with both system-level approaches and skill-level variants as mentioned in the above baselines.

\begin{table}[htbp]
\centering
\small
\begin{tabular}{lcc}
\toprule
Method & $Q(\mathcal H^t) \uparrow$ & $Hit@K$ $\uparrow$ \\
\midrule
HypoGeniC & 0.719 $\pm$ 0.014 & 0.472 $\pm$ 0.050 \\
HypotheSAEs & 0.582 $\pm$ 0.043 & 0.208 $\pm$ 0.038 \\
LLM-ZeroShot & 0.701 $\pm$ 0.006 & 0.516 $\pm$ 0.029 \\
LLM-FewShot & 0.719 $\pm$ 0.007 & 0.522 $\pm$ 0.022 \\
\midrule
No-Skill & 0.724 $\pm$ 0.040 & 0.409 $\pm$ 0.079 \\
AI-Generated-Skill & 0.738 $\pm$ 0.050 & 0.579 $\pm$ 0.022 \\
Human-Designed-Skill & 0.761 $\pm$ 0.040 & 0.497 $\pm$ 0.022 \\
\midrule
\textbf{{\name}} & \textbf{0.785 $\pm$ 0.030} & 
\textbf{0.648 $\pm$ 0.142} \\
\bottomrule
\end{tabular}
\caption{Hypothesis generation performance.}
\label{tab:generation_results}
\end{table}

As shown in Table~\ref{tab:generation_results}, {\name} achieves the best performance among all compared methods, obtaining the highest hypothesis quality score ($Q(\mathcal H^t)=0.785$) and Hit Rate ($0.648$). 
Compared with system-level baselines, {\name} improves Hit Rate over HypoGeniC and HypotheSAEs by 17.6\% and 44.0\%, respectively, demonstrating its ability to generate hypotheses with stronger scientific quality and broader coverage of reference hypotheses. 
It also consistently outperforms LLM-based prompting baselines, indicating that learned procedural skills provide more effective guidance than static instructions or few-shot demonstrations. 
For skill-level variants, removing skill learning from our framework decreases the Hit Rate from 0.648 to 0.409, while replacing learned skills with AI-generated or human-designed skills also leads to performance degradation. 
These results reflect that the proposed adversarial hypothesis generation framework effectively improves the overall performance of the hypothesis generation stage by introducing discriminator-guided refinement. 

\subsection{Hypothesis Testing Evaluation}
We then evaluate the hypothesis testing stage on test tasks using available ground-truth hypotheses. All methods are evaluated under the same experimental environment, where generated designs are executed on the provided datasets and assessed according to testing outcomes including experiment designs, experiment codes, and statistical results. 

\begin{table}[htbp]
\centering
\small
\begin{tabular}{lcc}
\toprule
Method & $T_h \uparrow$ & $E_h \uparrow$ \\
\midrule
POPPER & 0.417 $\pm$ 0.141 & \textbf{0.984 $\pm$ 0.000} \\
ReAct & 0.589 $\pm$ 0.078 & 0.859 $\pm$ 0.051 \\
LLM-ZeroShot & 0.516 $\pm$ 0.014 & 0.870 $\pm$ 0.030 \\
LLM-FewShot & 0.539 $\pm$ 0.008 & 0.966 $\pm$ 0.005 \\
\midrule
No-Skill & 0.562 $\pm$ 0.055 & 0.885 $\pm$ 0.099 \\
AI-Generated-Skill & 0.555 $\pm$ 0.090 & 0.891 $\pm$ 0.131 \\
Human-Designed-Skill & 0.612 $\pm$ 0.096 & 0.911 $\pm$ 0.043 \\
\midrule
\textbf{{\name}} & \textbf{0.659 $\pm$ 0.064} & 
\textbf{0.966 $\pm$ 0.016} \\
\bottomrule
\end{tabular}
\caption{Hypothesis testing performance on test tasks.}
\label{tab:testing_results}
\end{table}

As shown in Table~\ref{tab:testing_results}, {\name} achieves the highest test pass rate ($T_h=0.659$) while maintaining a competitive execution success rate ($E_h=0.966$). 
Compared with system-level baselines, {\name} substantially improves testing effectiveness over POPPER, ReAct, and LLM prompting methods, demonstrating the advantage of learning experiment design and execution strategies from previous testing trajectories. 
Notably, although POPPER achieves a higher execution success rate, its substantially lower test pass rate suggests that many of its generated experiments are technically executable but scientifically ineffective for validating the target hypotheses. 
Compared with skill-level variants, {\name} consistently outperforms No-Skill, AI-Generated-Skill, and Human-Designed-Skill, confirming that task-specific feedback-driven skill refinement is crucial for improving scientific testing capability.

\subsection{Skill Learning Analysis}
\begin{figure}[t]
    \centering
    \begin{subfigure}[b]{0.95\linewidth}
        \centering
        \includegraphics[width=\linewidth]{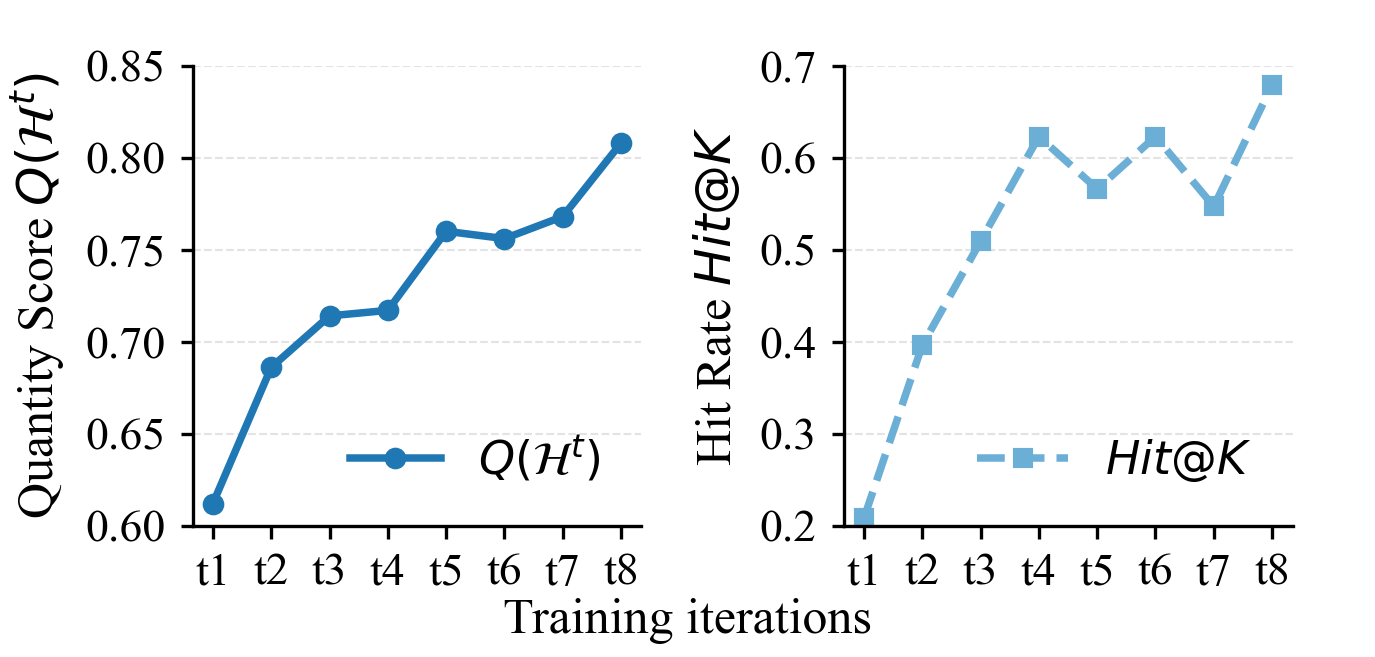}
        \caption{Hypothesis generation skill learning process.}
        \label{fig:skill_generation}
    \end{subfigure}
    
    \vspace{0.15cm}
    
    \begin{subfigure}[b]{0.95\linewidth}
        \centering
        \includegraphics[width=\linewidth]{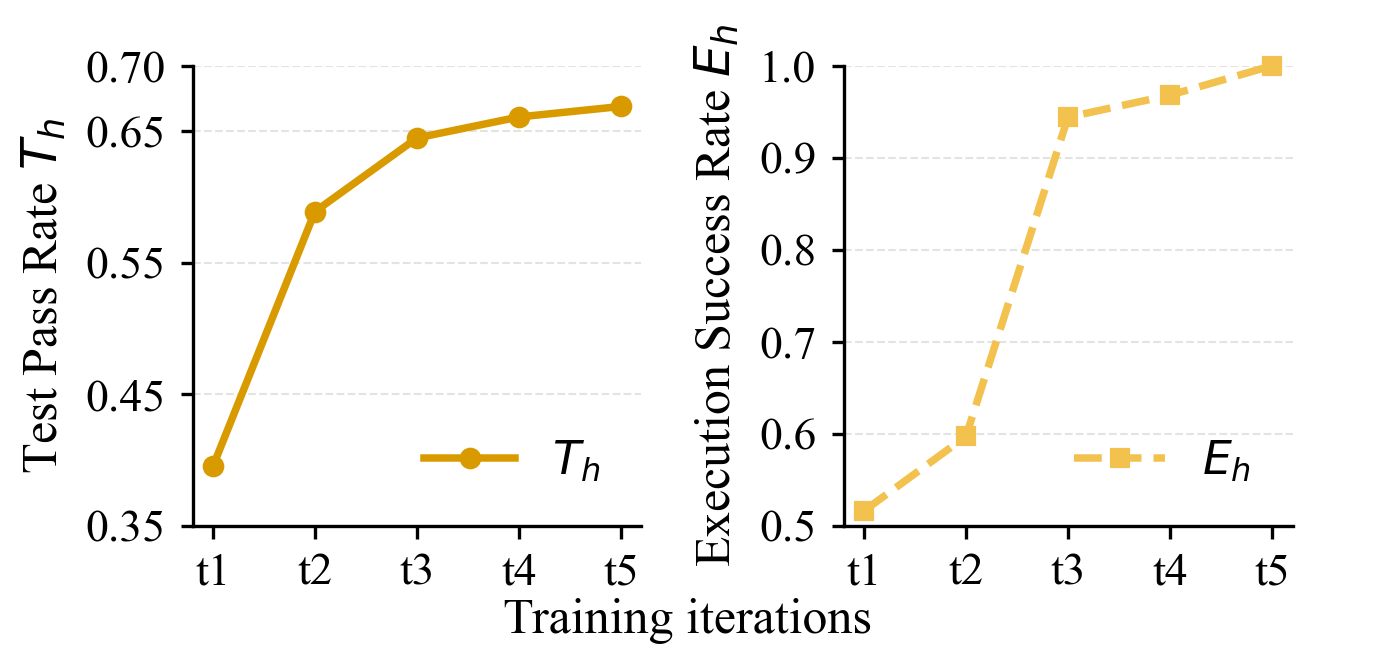}
        \caption{Hypothesis testing skill learning process.}
        \label{fig:skill_testing}
    \end{subfigure}
    
    \caption{Skill learning progress across refinement iterations. }
    \label{fig:skill_learning}
\end{figure}

To analyze how the proposed skill learning approach improves agent capabilities over iterations, we track the evolution of evaluation metrics during the skill refinement process. 
Figure~\ref{fig:skill_learning} shows the learning progress of hypothesis generation and testing skills.

For hypothesis generation, as shown in Figure~\ref{fig:skill_generation}, both $Q(\mathcal H^t)$ and Hit@K exhibit an overall increasing trend across refinement iterations. 
Specifically, $Q(\mathcal H^t)$ improves from 0.612 at the initial iteration to 0.808 after skill refinement, while Hit@K increases from 0.208 to 0.679, indicating that the learned skill enables the generator to produce hypotheses with not only higher scientific quality but also broader coverage of reference hypotheses. 
Although the metrics fluctuate in several intermediate iterations, the overall improvement demonstrates that discriminator feedback provides effective signals for refining hypothesis construction strategies rather than merely selecting high-scoring outputs.

For hypothesis testing, Figure~\ref{fig:skill_testing} shows consistent improvements in both testing effectiveness and execution correctness. 
The $T_h$ rate increases from 0.395 to 0.669, while $E_h$ improves from 0.516 to 1.000. 
This indicates that the learned testing skill progressively captures more effective experiment design and execution procedures.

\subsection{Ablation Study}
To evaluate the impact of refinement signals, we remove the corresponding supervision sources in both stages. 
For hypothesis generation, w/o Feedback removes discriminator feedback during skill refinement. For hypothesis testing, w/o Outcome removes testing results during skill refinement, while retaining the generated hypotheses, experiment designs, and executable codes.

As shown in Table~\ref{tab:ablation_feedback}, stage-specific refinement signals consistently improve performance. In hypothesis generation, the full model outperforms w/o Feedback in both $Q(\mathcal H^t)$ and $Hit@K$ (0.785 vs. 0.726 and 0.648 vs. 0.491), showing that discriminator feedback effectively refines generation skills. In hypothesis testing, removing execution outcomes decreases $T_h$ from 0.659 to 0.565 while maintaining comparable $E_h$, demonstrating the importance of empirical validation signals for improving testing skills.

\begin{table}[t]
\centering
\small
\begin{tabular}{lcc}
\toprule
Variant & $Q(\mathcal H^t) \uparrow$ & $Hit@K \uparrow$ \\
\midrule
w/o Feedback & 0.726 $\pm$ 0.008 & 0.491 $\pm$ 0.015 \\
Full & 0.785 $\pm$ 0.030 & 0.648 $\pm$ 0.142 \\
\bottomrule
\end{tabular}

\vspace{0.15cm}

\begin{tabular}{lcc}
\toprule
Variant & $T_h \uparrow$ & $E_h \uparrow$ \\
\midrule
w/o Outcome & 0.565 $\pm$ 0.084 & 0.957 $\pm$ 0.039 \\
Full & 0.659 $\pm$ 0.064 & 0.966 $\pm$ 0.016 \\
\bottomrule
\end{tabular}
\caption{Ablation study of skill learning.}
\label{tab:ablation_feedback}
\end{table}

%% file: Sections/05discussion.tex
\section{Discussion}

Our framework demonstrates the potential of experience-guided skill learning to improve the performance of hypothesis generation and testing without updating the underlying agents and foundation models. Despite these advantages, several limitations remain: (1) the effectiveness of skill refinement currently relies on the quantity and reliability of feedback information. Errors in hypothesis evaluation or insufficient ground-truth validation may lead to suboptimal skill updates. Future work could explore stronger scientific evaluators, multi-agent consensus mechanisms, or human-in-the-loop feedback to improve feedback reliability; (2) the current framework mainly focuses on data-driven scientific discovery with executable computational experiments. Extending skill learning to domains involving complex theoretical reasoning, wet lab, or interdisciplinary knowledge integration remains an important direction; 
(3)although learned skills may transfer across tasks, automatically determining the granularity, composition, and organization of scientific skills within a complex discovery pipeline remains an open challenge for achieving effective task performance.

%% file: Sections/06conclusion.tex
\section{Conclusion}

In this work, we propose \name, an experience-guided multi-agent framework for hypothesis generation and hypothesis testing in the scientific discovery process. Unlike existing AI scientist systems that treat discovery stages as independent tasks and rely on static model capabilities, \name enables AI agents to continuously improve their capabilities by learning procedural skills from previous discovery experiences. Specifically, we introduce adversarial refinement for hypothesis generation, where discriminator-guided feedback helps distill effective hypothesis construction strategies without explicit supervision, and experience-guided refinement for hypothesis testing, where empirical validation outcomes improve experiment design and execution skills. Extensive experiments on scientific discovery benchmarks demonstrate that \name consistently improves both hypothesis generation and testing performance, while the learned skills generalize across diverse scientific tasks. These results suggest a promising direction in autonomous scientific discovery: developing continually improving AI agent systems that accumulate experience and progressively strengthen their capabilities for scientific reasoning.